\documentclass[fleqn,usenatbib]{mnras}

\usepackage[T1]{fontenc}

\DeclareRobustCommand{\VAN}[3]{#2}
\let\VANthebibliography\thebibliography
\def\thebibliography{\DeclareRobustCommand{\VAN}[3]{##3}\VANthebibliography}

\usepackage[dvipdfmx]{graphicx}	
\usepackage{amsmath}	
\usepackage{amssymb}	
\usepackage[utf8x]{inputenc}
\usepackage[dvipdfmx]{color}
\usepackage{color}
\usepackage{bm}
\usepackage{color}
\usepackage{colortbl}
\usepackage{tikz}
\usetikzlibrary{trees}

\title[Scale analysis of angular momentum flux]{Scale-dependent analysis of angular momentum flux in high-resolution magnetohydrodynamic simulations for solar differential rotation}

\author[K. Mori and H. Hotta]{
K. Mori$^{1, 2}$\thanks{E-mail: mn@ras.org.uk (KTS)}
H. Hotta$^{2}$
\\
$^{1}$Department of Physics, Graduate School of Science, Chiba University 1-33 Yayoi-cho, Inage-ku, Chiba 263-8522 Japan\\
$^{2}$Institute for Space-Earth Environmental Research, Nagoya University, Furo-cho, Chikusa-ku, Nagoya, Aichi 464-8601, Japan
}

\date{Accepted XXX. Received YYY; in original form ZZZ}

\pubyear{2023}

\begin{document}
\label{firstpage}
\pagerange{\pageref{firstpage}--\pageref{lastpage}}
\maketitle

\begin{abstract}
In this work, we systematically investigate the scale-dependent angular momentum flux by analysing high-resolution three-dimensional magnetohydrodynamic simulations in which the solar-like differential rotation is reproduced without using any manipulations. More specifically, the magnetic angular momentum transport (AMT) plays a dominant role in the calculations. We examine the important spatial scales for the magnetic AMT. The main conclusions of our approach can be summarized as follows: 1. Turbulence transports the angular momentum radially inward. This effect is more pronounced in the highest resolution calculation. 2. The dominant scale for the magnetic AMT is the smallest spatial scale. 3. The dimensionless magnetic correlation is low in the high-resolution simulation. Thus, chaotic but strong small-scale magnetic fields achieve efficient magnetic AMT.
\end{abstract}




\section{Introduction}
\label{sec:introduction}
The Sun's differential rotation (DR) is a crucial process for generating its magnetic field through the $\Omega$ effect \citep{parker_1955ApJ...122..293P}.
The existence of the DR effect was experimentally demonstrated in the early 1630s by tracking sunspots \citep{paterno_2010Ap&SS.328..269P}.
In the modern era, helioseismology has revealed the detailed profile of the DR \citep[e.g.,][]{schou_1998ApJ...505..390S}. The fast equator is considered to be the most prominent feature of the solar DR. In particular, the solar equator rotates in a 25-day period, while the polar region rotates in a 30-day period.\par
From a theoretical (numerical) viewpoint, solar-like DR, where the equator rotates faster than the polar regions, can be reproduced with a low Rossby number (Ro) (i.e. strong rotational influence), where the Ro is defined with $\mathrm{Ro}=v_\mathrm{c}/(2\Omega_0 L).$ $v_\mathrm{c}$, $\Omega_0$, and $L$ refer to the typical convection velocity, the system rotation rate, and the typical convection spatial scale, respectively. When the system is located in a high-Ro regime (i.e. weak rotational influence), the polar region starts to rotate faster, which is called the anti-solar DR \citep[e.g.,][]{gastine_2014MNRAS.438L..76G}. There is a consensus in the scientific community that the Sun lies in the low-Ro regime because it has a solar-like DR.
However, the convective conundrum related to solar angular momentum transport (AMT) has raised questions about the former assumption. Alarmingly, high-resolution simulations, even those including solar parameters, such as solar rotation rate and luminosity, can easily fail to simulate the solar-like DR \citep{hotta_2015ApJ...798...51H,omara_2016AdSpR..58.1475O}. Moreover, the high-resolution simulation tends to generate a high-Ro regime, which accelerates the polar region \cite[see][hereafter MH23]{mori_2023MNRAS.519.3091M}\par
Recently, \cite{hotta_2021NatAs...5.1100H} (hereafter HK21) solved this problem. When the resolution was increased to unprecedentedly high levels, the magnetic energy exceeded the kinetic energy. As a result, the Maxwell stress can transport the angular momentum and lead to a fast equator \cite[see also][hereafter HKS22]{hotta_2022ApJ...933..199H}.
According to the literature, the equatorial acceleration can be achieved by outward AMT from the rotational axis with rotationally constrained turbulence \citep[][MH23]{karak_2015A&A...576A..26K,kapyla_2022arXiv220700302K}.
In their high-resolution simulation, however, HKS22 found that the magnetic field is the dominant component of the AMT.
Even though the scale analysis by HKS22 implies that small-scale magnetic fields can be the dominant aspect of AMT, HKS22 did not decompose the angular momentum flux (AMF) into separate scales.
Thus, HKS22 did not identify any direct evidence for the importance of the small-scale magnetic fields in the AMT. Thus we analyse the HK21 calculation results using the spatial scale decomposition method for the AMF suggested in MH23.
We address three main issues in this work:
\begin{itemize}
	\item[A.]What is the most important scale of the magnetic field for accelerating the equator?
	\item[B. ]Does the inward AMT by the turbulence become stronger at smaller scales?
	\item[C. ]Is the magnetic field important because the magnetic field is strong or because the magnetic fields are correlated?
\end{itemize}
We thoroughly investigate the importance of the small-scale magnetic field for the AMT using the above scale decomposition (Issue A). MH23 suggested that the smaller scale turbulence has stronger radially inward AMT with a weaker rotational influence. We can confirm this pattern by analysing high-resolution simulations (Issue B). HKS22 found also that the magnetic AMT is dominant in their high-resolution calculation. Since the momentum flux is a form of covariance $\propto\bm{BB}$, the strong magnetic AMT can be interpreted by considering two possibilities. The first one is related to the fact that just the magnetic field strength $|B|$ is high. MH21 already reported that the magnetic field is significantly strong in their calculation, which could explain this effect. The other possibility is that the dimensionless correlation $\overline{\bm{BB}}$ is high. In particular, HKS22 found that the correlation between the magnetic fields is generated by high-Ro turbulence. Typically, turbulence on a smaller scale is less affected by the rotational influence; that is, the higher Ro. Thus, smaller-scale magnetic fields may be more strongly correlated than larger-scale magnetic fields (Issue C).
\par
This manuscript is constructed as follows. We describe the model setup for the numerical simulations in Section \ref{sec:model}. We present the proposed method to decompose the AMT in Section \ref{sec:analysis}. Next, we thoroughly discuss the results of our analysis in Section \ref{sec:result}. The scale decomposition of the AMF and the correlation are also shown in Section \ref{sec:result}. Finally, we summarize and conclude the paper in Section \ref{sec:conclusion}.

\section{Numerical Model}
\label{sec:model}
We explain our developed numerical model setup in this section. We analyse the results of the HK21 calculations, which provide three different resolution calculations: namely, Low, Middle, and High (see Table \ref{ta:resol}).
\begin{table}
  \centering
  \caption{Number of grid points in each case. For our analysis, we convert the Yin--Yang grid into an ordinary spherical grid using the grid points of $N_r \times 2N_\theta \times 4N_\phi/3$.}
  \label{ta:resol}
  \begin{tabular}{lc}
    \hline
    Case & No. of Grid points\\
	& $N_r\times N_\theta\times N_\phi\times 2$ \\
    \hline
	Low &  $96\times384\times1152$ \\
	Middle & $192\times768\times2304$\\
	High & $384\times1536\times4608$ \\
    \hline
  \end{tabular}
\end{table}
HKS22 supplied dimensionless quantities for the calculations (see their Table 1). We convert the Yin--Yang grid \citep{kageyama_2004GGG.....5.9005K} into spherical geometry for analyses.
The radial extent of the computational domain is $0.71R_{\odot} \leq r \leq 0.96R_{\odot}$, where $R_\odot$ is the solar radius. We use the R2D2 (Radiation and RSST for Deep Dynamics) code \citep{hotta_2019SciA....5.2307H,hotta_2020MNRAS.494.2523H,hotta_2021NatAs...5.1100H}, where RSST represents the reduced speed of sound technique \citep{hotta_2012A&A...539A..30H}. We use Model S \citep{christensen_1996Sci...272.1286C} for the background stratification and related variables. We also use the solar luminosity $L_\odot$ and solar rotation rate $\Omega_\odot$ (see HKS22 for more details).
The calculation continues for $4000$ days for each case and the following results are averaged between $t=3600$ to $4000$ days unless otherwise noted.

\section{Decomposition method}
\label{sec:analysis}
In this work, we decompose the Reynolds (flow) and Maxwell (magnetic) AMFs\footnote{The AMF caused by the Reynolds (Maxwell) stress is called the ``Reynolds (Maxwell) AMF'' in this study.} to scale-dependent values using the Fourier transforms in the longitudinal direction. MH23 suggested using the spatial scale decomposition method. The Reynolds $F_{\mathrm{R}, \alpha}$ and Maxwell $F_{\mathrm{M}, \alpha}$ AMFs are written as follows:
\begin{align}
	&F_{\mathrm{R},\alpha} = \rho_0\lambda\langle v_\alpha'v_\phi'\rangle, & \nonumber \\
	&F_{\mathrm{M},\alpha} = -\lambda\frac{\langle B_\alpha B_\phi\rangle}{4\pi}, &
\end{align}
where $\alpha = r$ or $\theta$, $\lambda = r\sin\theta$, and $v$ is the fluid velocity in the rotating frame.
The parenthesis $\langle\rangle$ denotes the longitudinal average.
$'$ represents perturbations from the longitudinal average, such as $\bm{v}' = \bm{v} - \langle\bm{v}\rangle$.
We decompose these AMFs using Parseval's theorem as follows:
\begin{align}
	&F_{\mathrm{R},\alpha} = \rho_0\lambda \langle v'_\alpha v'_\phi\rangle = \sum_{i=1}^4 f^i_{\mathrm{R},\alpha},\nonumber \\
	&F_{\mathrm{M},\alpha} = -\lambda \frac{\langle B_\alpha B_\phi\rangle}{4\pi} \simeq -\lambda \frac{\langle B'_\alpha B'_\phi\rangle}{4\pi} = \sum_{i=1}^4 f^i_{\mathrm{M},\alpha},
\end{align}
The mean magnetic field $\langle\bm{B}\rangle$ is weak compared with the perturbation $\bm{B}'$; therefore, we exclude the contribution by the $-\lambda\langle B_\alpha\rangle\langle B_\phi\rangle/4\pi$.
We do not decompose the scale based on the wavenumber $m$, but on the actual spatial scale (see MH23, particularly eqs (17)--(20) and Table 1). The scale-dependent AMF $f_{\mathrm{M},\alpha}^i$ can be defined as follows:
\begin{align}
	f^i_{\mathrm{M},\alpha}(r,\theta) =
	-\frac{\lambda}{2\pi}\sum_{m=m_{i\mathrm{(min)}}}^{m_{i\mathrm{(max)}}} \mathrm{Re}
  \left[\widehat{B}_\alpha \widehat{B}^*_\phi\right],
\end{align}
where $\widehat{\ \ }$ refers to the Fourier transform in the longitudinal direction defined by MH23 (see their eq. (12)) and $^*$ indicates the complex conjugate.
When $m_{i(\mathrm{max})}=N_\phi/2$, the expression becomes:
\begin{align}
  f^i_{\mathrm{M},\alpha}(r,\theta) = -\frac{\lambda}{4\pi}  \left(
  2\sum_{m=m_{i\mathrm{(min)}}}^{m_{i\mathrm{(max)}}-1}\mathrm{Re}
  \left[\widehat{B}_\alpha \widehat{B}^*_\phi\right]  \right. \nonumber\\
  \left.+\mathrm{Re}\left[\widehat{B}_\alpha\left(\frac{N_\phi}{2}\right)\widehat{B}_\phi\left(\frac{N_\phi}{2}\right)\right]
  \right),
\end{align}
where $N_\phi$ is the number of grid points in the longitudinal direction. $m_{i\mathrm{(min)}}$ and $m_{i\mathrm{(max)}}$ are calculated based on the maximum $L_{i\mathrm{(max)}}$ and minimum $L_{i\mathrm{(min)}}$ scale length for the scale-dependent AMF $f^i_{\mathrm{R,~\alpha}}$ and $f^i_{\mathrm{M,~\alpha}}$ as:
\begin{align}
  m_{i\mathrm{(min)}}(r,\theta) &= \mathrm{floor}\left(\frac{2\pi\lambda}{L_{i\mathrm{(max)}}}\right) + 1, \\
  m_{i\mathrm{(max)}}(r,\theta) &= \mathrm{floor}\left(\frac{2\pi\lambda}{L_{i\mathrm{(min)}}}\right),
\end{align}
where $\mathrm{floor}()$ is the floor function. $L_{i\mathrm{(max)}}$ and $L_{i\mathrm{(min)}}$ in this study is presented in Table \ref{ta:l}, which is the same as that in MH23.
\begin{table}
  \centering
  \caption{$L_{i\mathrm{(max)}}$ and $L_{i\mathrm{(min)}}$ to determine $m_{i\mathrm{(min)}}$ and $m_{i\mathrm{(max)}}$ and resulting $f^i_{\mathrm{M},\alpha}$ is shown. $R_{\odot}$ is the solar radius.}
  \label{ta:l}
  \begin{tabular}{lcc}
    \hline
    $i$ & $L_{i\mathrm{(min)}}$ & $L_{i\mathrm{(max)}}$ \\
    \hline
    1 & 240~Mm & $2\pi R_{\odot}$ \\
    2 & 120~Mm & 240~Mm\\
    3 & 60~Mm  & 120~Mm\\
    4 & $4\pi\lambda/N_\phi $ & 60~Mm \\
    \hline
  \end{tabular}
\end{table}
Our definition of the scale-dependent AMF covers the scales $L_{i\mathrm{(min)}}\leq L_m< L_{i\mathrm{(max)}}$ (see MH23 for more details).
We also define the dimensionless correlation as follows:
\begin{align}
	&\tilde{F}_{\mathrm{R},\alpha} \equiv \frac{\langle v_\alpha'v_\phi'\rangle}{[v_\alpha]_\mathrm{RMS}[v_\phi]_\mathrm{RMS}},& \nonumber \\
	&\tilde{F}_{\mathrm{M},\alpha} \equiv -\frac{\langle B_\alpha'B_\phi'\rangle}{[B_\alpha]_\mathrm{RMS}[B_\phi]_\mathrm{RMS}}.
\end{align}
where $[]_{\mathrm{RMS}}$ is the root-mean-square (RMS) in the longitudinal direction.
$\tilde{F}_{\mathrm{R},\alpha}$ and $\tilde{F}_{\mathrm{M},\alpha}$ are useful because they demonstrate the dominant contribution of the AMF, that is, the dimensionless correlations or the amplitude of the physical quantity.
We also define the dimensionless scale-dependent correlation $\tilde{f}_{\mathrm{R},\alpha}$ and $\tilde{f}_{\mathrm{M},\alpha}$ as follows:
\begin{align}
	&\tilde{f}_{\mathrm{R},\alpha} = \frac{\sum_{m=m_{i\mathrm{(min)}}}^{m_{i\mathrm{(max)}}} \mathrm{Re}
  \left[\widehat{v}_\alpha \widehat{v}^*_\phi\right]}{\left(\sqrt{\sum_{m=m_{i\mathrm{(min)}}}^{m_{i\mathrm{(max)}}}|\widehat{v}_\alpha|^2}\right)\left(\sqrt{\sum_{m=m_{i\mathrm{(min)}}}^{m_{i\mathrm{(max)}}}|\widehat{v}_\phi|^2}\right)},& \nonumber \\
	&\tilde{f}_{\mathrm{M},\alpha} = \frac{-\sum_{m=m_{i\mathrm{(min)}}}^{m_{i\mathrm{(max)}}} \mathrm{Re}
  \left[\widehat{B}_\alpha \widehat{B}^*_\phi\right]}{\left(\sqrt{\sum_{m=m_{i\mathrm{(min)}}}^{m_{i\mathrm{(max)}}}|\widehat{B}_\alpha|^2}\right)\left(\sqrt{\sum_{m=m_{i\mathrm{(min)}}}^{m_{i\mathrm{(max)}}}|\widehat{B}_\phi|^2}\right)}.&
\end{align}
When $m_{i\mathrm{(max)}} = N_\phi/2$, the expressions become
\begin{align}
	\tilde{f}_{\mathrm{R},\alpha} = &\left(2\sum_{m=m_{i\mathrm{(min)}}}^{m_{i\mathrm{(max)}}-1} \mathrm{Re}
  \left[\widehat{v}_\alpha \widehat{v}^*_\phi\right]+\mathrm{Re}\left[\widehat{v}_\alpha\left(\frac{N_\phi}{2}\right)\widehat{v}_\phi\left(\frac{N_\phi}{2}\right)\right]\right)& \nonumber \\
	&\left/\left(\sqrt{2\sum_{m=m_{i\mathrm{(min)}}}^{m_{i\mathrm{(max)}}-1}|\widehat{v}_\alpha|^2 + \left|\widehat{v}_{\alpha}\left(\frac{N_\phi}{2}\right)\right|^2}\right)\right.&\nonumber \\
	&\left/\left(\sqrt{2\sum_{m=m_{i\mathrm{(min)}}}^{m_{i\mathrm{(max)}}-1}|\widehat{v}_\phi|^2 + \left|\widehat{v}_{\phi}\left(\frac{N_\phi}{2}\right)\right|^2}\right)\right.,&\nonumber \\
	\tilde{f}_{\mathrm{M},\alpha} = &\left(2\sum_{m=m_{i\mathrm{(min)}}}^{m_{i\mathrm{(max)}}-1} \mathrm{Re}
  \left[\widehat{B}_\alpha \widehat{B}^*_\phi\right]+\mathrm{Re}\left[\widehat{B}_\alpha\left(\frac{N_\phi}{2}\right)\widehat{B}_\phi\left(\frac{N_\phi}{2}\right)\right]\right)& \nonumber \\
	&\left/\left(\sqrt{2\sum_{m=m_{i\mathrm{(min)}}}^{m_{i\mathrm{(max)}}-1}|\widehat{B}_\alpha|^2 + \left|\widehat{B}_{\alpha}\left(\frac{N_\phi}{2}\right)\right|^2}\right)\right.&\nonumber \\
	&\left/\left(\sqrt{2\sum_{m=m_{i\mathrm{(min)}}}^{m_{i\mathrm{(max)}}-1}|\widehat{B}_\phi|^2 + \left|\widehat{B}_{\phi}\left(\frac{N_\phi}{2}\right)\right|^2}\right)\right.\times (-1).&\nonumber \\
\end{align}

\section{RESULTS}
\label{sec:result}
\subsection{General properties}
\begin{figure}
	\begin{center}			
		\includegraphics[clip, width=7cm]{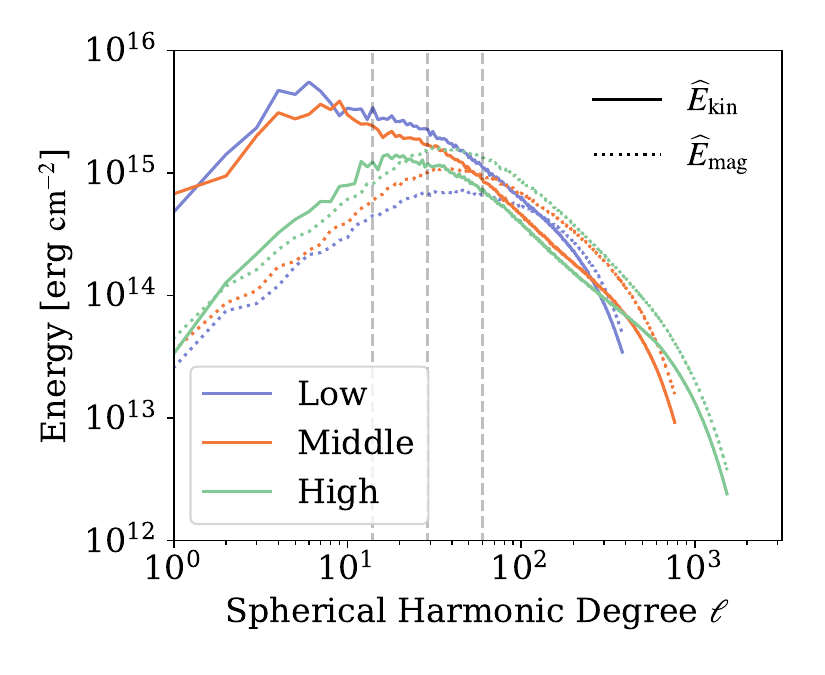}
    \caption{The kinetic energy $E_{\mathrm{k}}$ and magnetic energy $E_{\mathrm{m}}$ spectra at $r=0.85R_{\odot}$ are shown (A similar plot is shown in Fig. 9 of HKS22). The solid and dotted lines display $E_{\mathrm{k}}$ and $E_{\mathrm{m}}$, respectively. In this plot, only the $m \neq 0$ mode is shown to exclude contributions from the DR. The blue, orange, and green lines represent the results from the Low, Middle, and High cases, respectively. The large-scale kinetic energy ($\ell<30$) is significantly reduced in the High case. The three vertical dashed lines indicate $\ell$ corresponding to $240$, $120$, and $60\ \mathrm{Mm}$, respectively, which divide the scale-dependent AMFs $f^i_{\mathrm{R},\alpha}$ and $f^i_{\mathrm{M}, \alpha}$.}
	\label{spectrum}
	\end{center}
\end{figure}

Fig. \ref{spectrum} shows the kinetic and magnetic energy spectra (see eqs (16) and (17) of HKS22 for the definition). The blue, orange, and green lines show the results from the Low, Middle, and High cases, respectively.
As shown in HK21, the large-scale kinetic energy ($\ell<30$) is significantly reduced in the High case, whereas the magnetic energy $E_{\mathrm{m}}$ increases as resolution becomes higher in all scales.
See Fig. 10 in HKS22 for the DR and the meridional flow.
The Low case has anti-solar DR, whereas the High case possesses solar-type DR.

\begin{figure}
	\begin{center}
		\includegraphics[width = 0.5\textwidth]{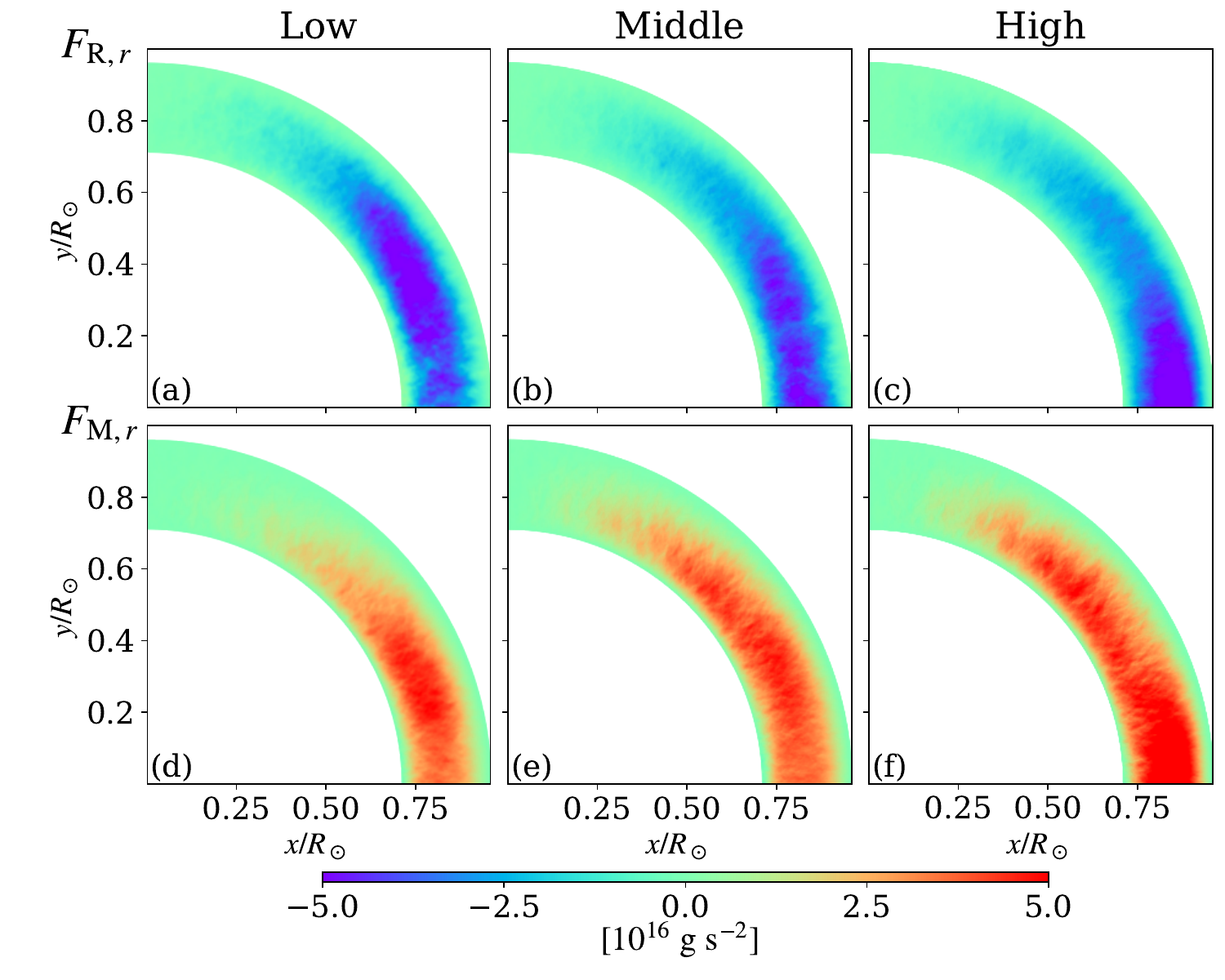}
    \caption{The radial Reynolds $F_{\mathrm{R},r}=\rho_0\lambda\langle v_{r}'v_{\phi}'\rangle$ and Maxwell $F_{\mathrm{M},r} = -\lambda\langle B_\mathrm{r}B_\phi\rangle/4\pi$ AMFs are shown.
		The upper and lower panels show $F_{\mathrm{R},r}$, and $F_{\mathrm{M},r}$, respectively.
		The result from Low (panels a and d), Middle (panels b and e), and High (panels c and f) cases are shown.
    }
		\label{flux_all}
	\end{center}
\end{figure}

Fig. \ref{flux_all} shows the radial Reynolds $F_{\mathrm{R},r}$ and Maxwell $F_{\mathrm{M},r}$ AMF. Note that almost the same figure is presented by HKS22 (see their Fig. 29). Here, we review their results.
The radial AMF significantly depends on the resolution; therefore, our main focus in this work is only on the radial AMFs. The radial Reynolds AMFs are always negative (i.e., inward AMTs) and their strengths are almost the same (Figs \ref{flux_all}a, b, and c). 
The radial Maxwell AMFs show positive values, that is, the outward AMT in all cases (Figs \ref{flux_all}d, e and f). The flux strength increases as the resolution increases.
\par
\begin{figure}
	\begin{center}
		\includegraphics[width = 0.5\textwidth]{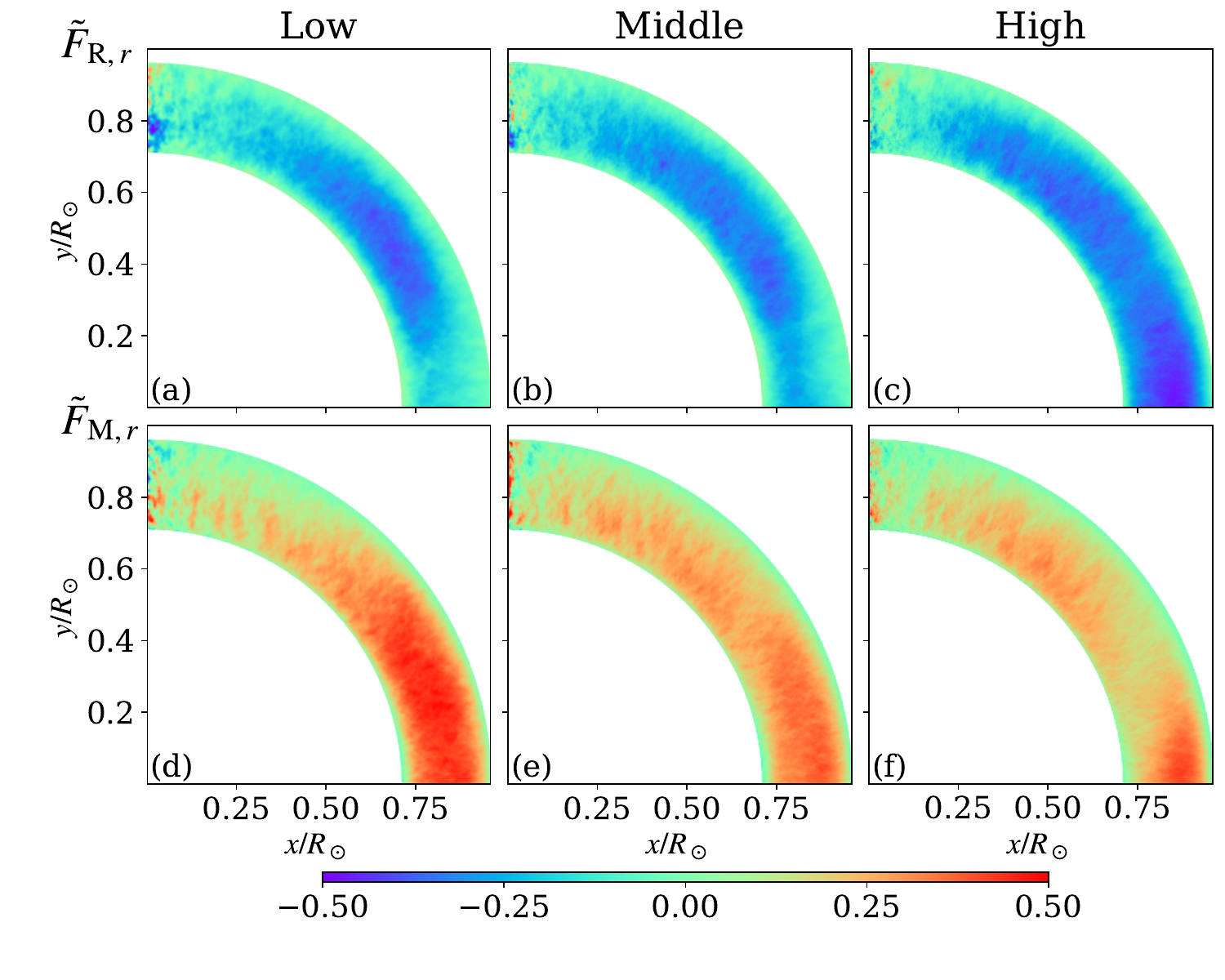}
    \caption{The dimensionless velocity $\tilde{F}_{\mathrm{R},r}$ and magnetic $\tilde{F}_{\mathrm{M},r}$ correlations are shown. The upper and lower panels show $\tilde{F}_{\mathrm{R},r}$ and  $\tilde{F}_{\mathrm{M},r}$, respectively.
		The result from Low (panels a and d), Middle (panels b and e), and High (panels c and f) cases are shown.
    }
		\label{normflux_all}
	\end{center}
\end{figure}

Fig. \ref{normflux_all} shows the dimensionless velocity $\tilde{F}_{\mathrm{R},r}$ and magnetic $\tilde{F}_{\mathrm{M},r}$ correlations. We also note that $\tilde{F}_{\mathrm{R},r}$ is already shown in HKS22 (see their Fig. 36), but $\tilde{F}_{\mathrm{M},r}$ is the original result in this work.
We can see the strongest negative correlation in the High case (Fig. \ref{normflux_all}c). This may be due to the high Ro in the High case (see Table 1 of HKS22 for Local Rossby number $\mathrm{Ro}_\ell$).
When the Ro is high, the downflow $v_r < 0$ becomes dominant. This downflow is bent by the Coriolis force and turbulence has negative correlations ($\langle v_r'v_\phi'\rangle < 0$).
This result indicates that $\tilde{F}_{\mathrm{R},r}$ makes a significant contribution to the radially inward Reynolds AMT $F_{\mathrm{R},r}$.
This strong negative correlation is thought to be caused by the high Ro due to the small-scale turbulence reproduced by the high resolution. Considering the magnetic field, the positive correlation decreases with increased resolution (Fig. \ref{normflux_all}d, e, and f).
This result indicates that the strength of the magnetic field is responsible for the radially outward AMT and the resulting fast equator (Issue C in Section \ref{sec:introduction}).
\subsection{Scale-dependent angular momentum flux}
In this subsection, we investigate the scale-dependent AMF.
As explained in section \ref{sec:analysis}, we decompose the AMFs into four scales (see also Table \ref{ta:l}).
We note that the index $i$ of $f_{\mathrm{R}, r}^i$ and $f_{\mathrm{M},r}^i$ does not represent the wavenumber $m$ (see Section \ref{sec:analysis} for details). \par
\begin{figure}
	\begin{center}
		\includegraphics[width = 0.5\textwidth]{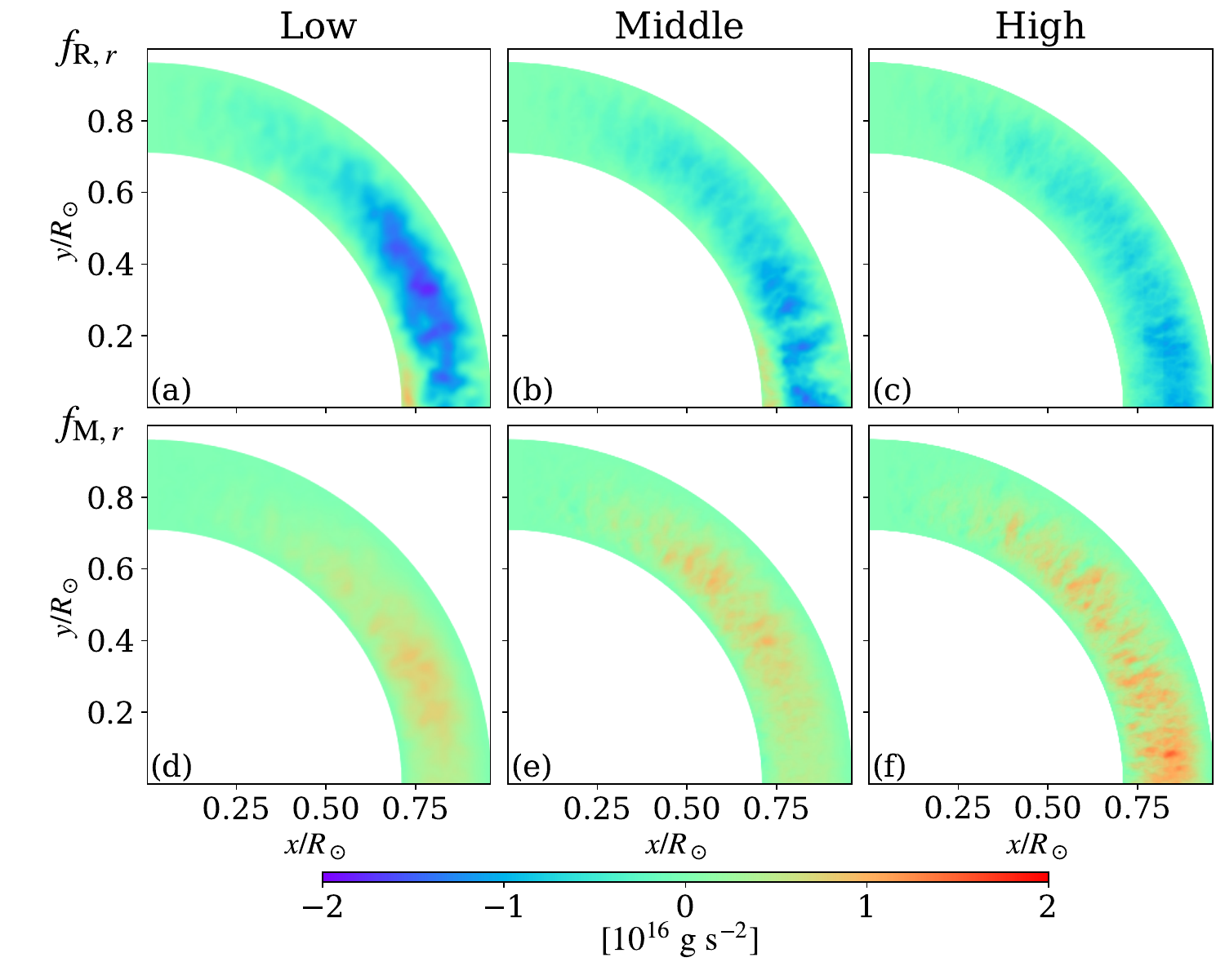}
    \caption{Depiction of the scale-dependent AMF $f^1_{\mathrm{R},r}$ and $f^1_{\mathrm{M},r}$, which correspond to the scale of $L_m\ge240~\mathrm{Mm}$. The format of the panels is identical to Fig. \ref{flux_all}.
    A Gaussian filter with five grid points width is also applied in all directions to reduce the realization noise.}
	\label{flux4_1}
	\end{center}
\end{figure}

Fig. \ref{flux4_1} shows the $f^1_{\mathrm{R},r}$ and $f^1_{\mathrm{M},r}$, which correspond to the spatial scale of $L_m\ge240~\mathrm{Mm}$, that is, the largest scale.
We use the same colour scale in Fig. \ref{flux4_1} and the following two figures for comparison.
We extract negative Reynolds AMFs $f^1_{\mathrm{R},r}$ in all cases (Fig. \ref{flux4_1}a, b, and c).
We find that the inward Reynolds transport is largest in the Low case because of its larger convection velocity (Fig. \ref{spectrum}). The Maxwell AMFs $f^1_{\mathrm{M},r}$ are positive in all cases (Fig. \ref{flux4_1}d, e, and f). The outward transport is largest in the High case and the magnetic field strength is highest in the High case, which probably leads to the strong outward AMT (see Fig. \ref{spectrum}).\par
\begin{figure}
	\begin{center}
		\includegraphics[width = 0.5\textwidth]{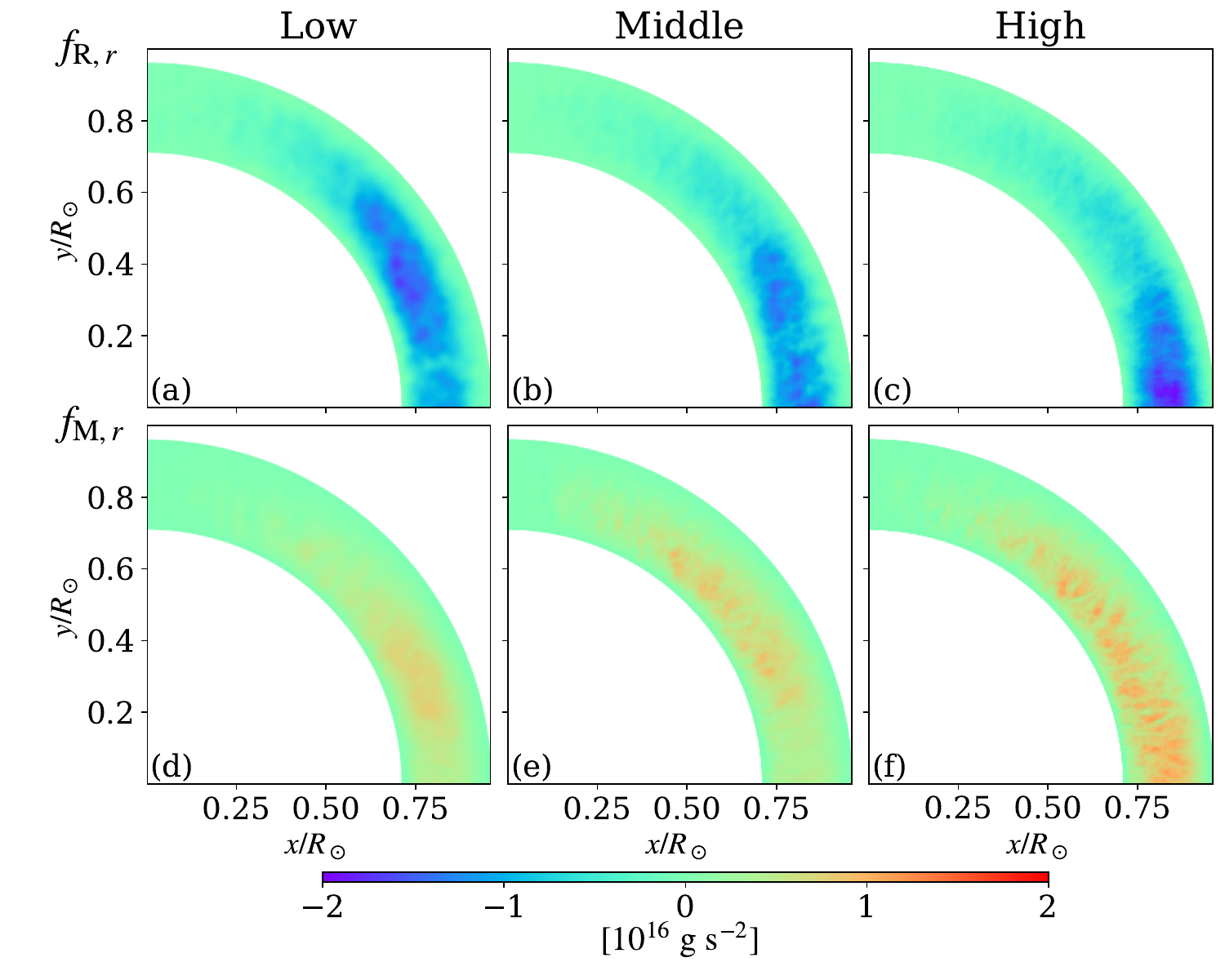}
    \caption{The same figure as Fig. \ref{flux4_1}, but for $f^2_{\mathrm{R},r}$ and $f^2_{\mathrm{M},r}$, which correspond to the scale of $120~\mathrm{Mm}\le L_m<240~\mathrm{Mm}$ is shown.}
	\label{flux4_2}
	\end{center}
\end{figure}

Fig. \ref{flux4_2} shows the $f^2_{\mathrm{R},r}$ and the $f^2_{\mathrm{M},r}$, which correspond to the spatial scale of $120~\mathrm{Mm}\le L_m<240~\mathrm{Mm}$.
Similar to $i=1$, the Reynolds $f^2_{\mathrm{R},r}$ and Maxwell $f^2_{\mathrm{M},r}$ are negative and positive in all cases, respectively. We can observe the stronger Reynolds AMFs (i.e., $|f^1_{\mathrm{R},r}| < |f^2_\mathrm{R,r}|$), especially in the Middle and High cases, while the kinetic energy is larger in $L_m>240~\mathrm{Mm}$ (Fig. \ref{spectrum}).
We can see the similar Maxwell AMFs, i.e.,  $f^1_{\mathrm{M},r} \simeq f^2_{\mathrm{M},r}$ in all cases. \par
 $f^3_{\mathrm{R},r}$ and $f^3_{\mathrm{M},r}$ are almost the same as $f^2_{\mathrm{R},r}$ and $f^2_{\mathrm{M},r}$, respectively, and are not shown here.\par
\begin{figure}
	\begin{center}
		\includegraphics[width = 0.5\textwidth]{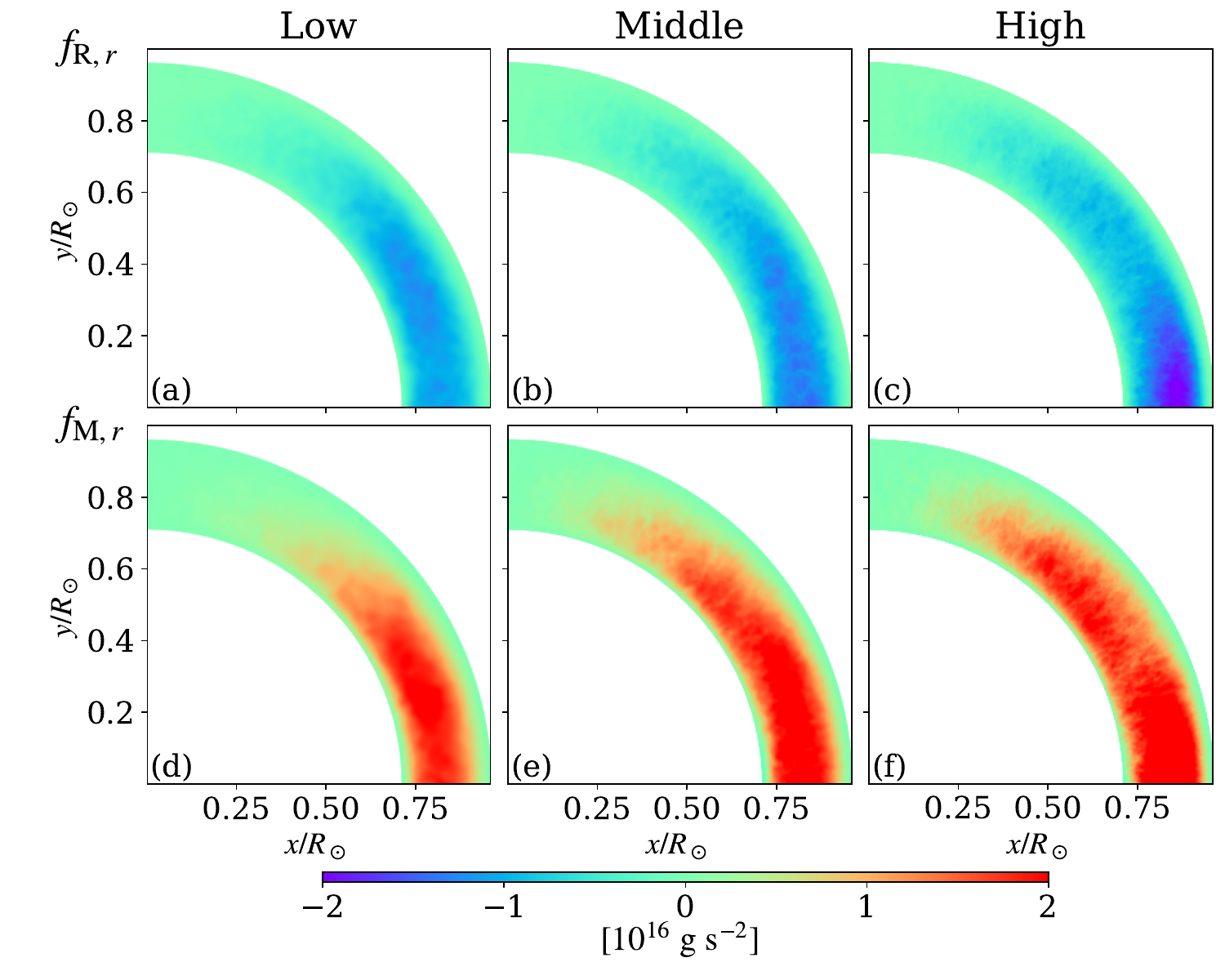}
    \caption{The same figure as Fig. \ref{flux4_1}, but for $f^4_{\mathrm{R},r}$ and $f^4_{\mathrm{M},r}$, which correspond to the scale of $L_m<60~\mathrm{Mm}$.
		We can find the strongest inward Reynolds and outward Maxwell AMFs in this scale.}
	\label{flux4_4}
	\end{center}
\end{figure}

$f_{\mathrm{R}, r}^{4}$ and $f^4_{\mathrm{M},r}$ are shown in Fig. \ref{flux4_4}.
The corresponding spatial scale is $L_m<60~\mathrm{Mm}$, which is the smallest scale.
The Reynolds AMFs $f^4_{\mathrm{R},r}$ are negative in all cases. More specifically, in the High case, the inward AMT near the equator is stronger than those at other scales (Issue B in Section \ref{sec:introduction}). The Maxwell AMFs $f^4_{\mathrm{M},r}$ are positive in all cases. We record the largest amplitude in this scale compared with the other scales, especially in the High case. Thus, we can argue that the small-scale ($<60~\mathrm{Mm}$) magnetic field is the dominant source of the AMT, as expected in HKS22 (Issue A in Section \ref{sec:introduction}).

\subsection{Scale-dependent dimensionless correlation}
In this subsection, we use the dimensionless correlation to investigate the physical origin of the AMFs, namely, the dimensionless correlation or spectral amplitude.\par
\begin{figure}
	\begin{center}
		\includegraphics[width = 0.5\textwidth]{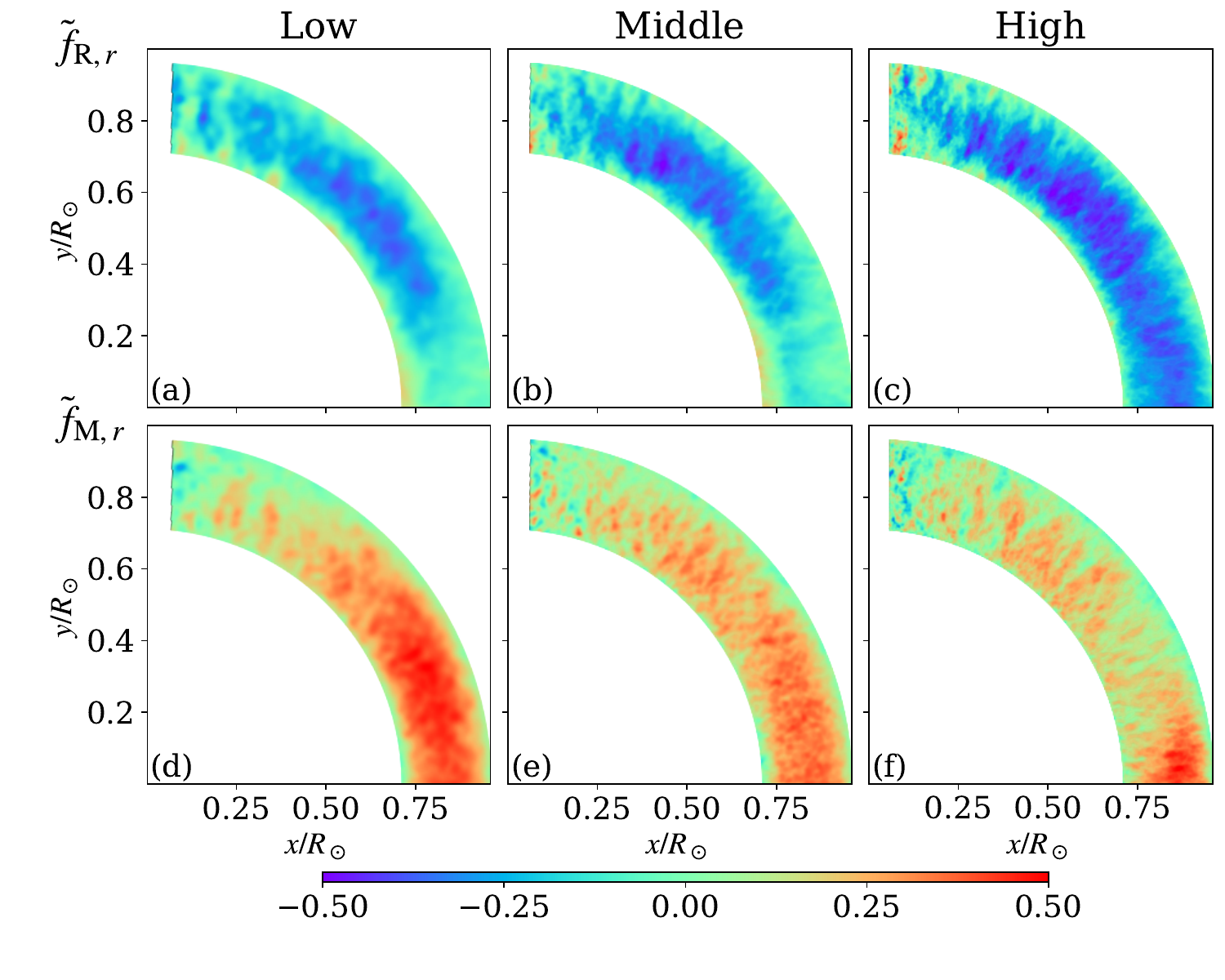}
    \caption{Depiction of scale-dependent dimensionless correlations $\tilde{f}^1_{\mathrm{R},r}$ and $\tilde{f}^1_{\mathrm{M},r}$, which correspond to the scale of $L_m\ge240~\mathrm{Mm}$. The format of the panels is identical to Fig. \ref{normflux_all}.
    A Gaussian filter with a five-grid-point width is also applied in all directions to reduce the realization noise.}
	\label{normflux4_1}
	\end{center}
\end{figure}

Fig. \ref{normflux4_1} shows the $\tilde{f}^1_{\mathrm{R},r}$ and $\tilde{f}^1_{\mathrm{M},r}$, which correspond to the spatial scale of $L_m\ge240~\mathrm{Mm}$.
The absolute value of the negative velocity correlation $\tilde{f}^1_{\mathrm{R},r}$, which is related to the inward AMT, is largest in the High case. The stronger correlation is likely due to the high $\mathrm{Ro}_{\ell}$ in the High case (see Table 1 of HKS22). The $\mathrm{Ro}_{\ell}$ can be derived from the calculation results before the scale decomposition process. It is interesting to note that we can still see the difference between resolutions; that is, the small-scale influence in the High case, as well as the after-scale decomposition. Figs \ref{flux4_1}, and \ref{normflux4_1} illustrate that the large velocities are responsible for the large Reynolds AMF in the Low case at this scale. While the correlation is large in the High case, the small velocity amplitude leads to a small Reynolds AMF. The absolute value of positive magnetic correlation $\tilde{f}^1_{\mathrm{M},r}$ is the largest in the Low case. The higher-resolution simulation can reproduce small-scale random motions, the correlation is smaller in the High case. This effect is also in direct line with the scale-integrated AMF (Fig. \ref{flux_all}).\par
\begin{figure}
	\begin{center}
		\includegraphics[width = 0.5\textwidth]{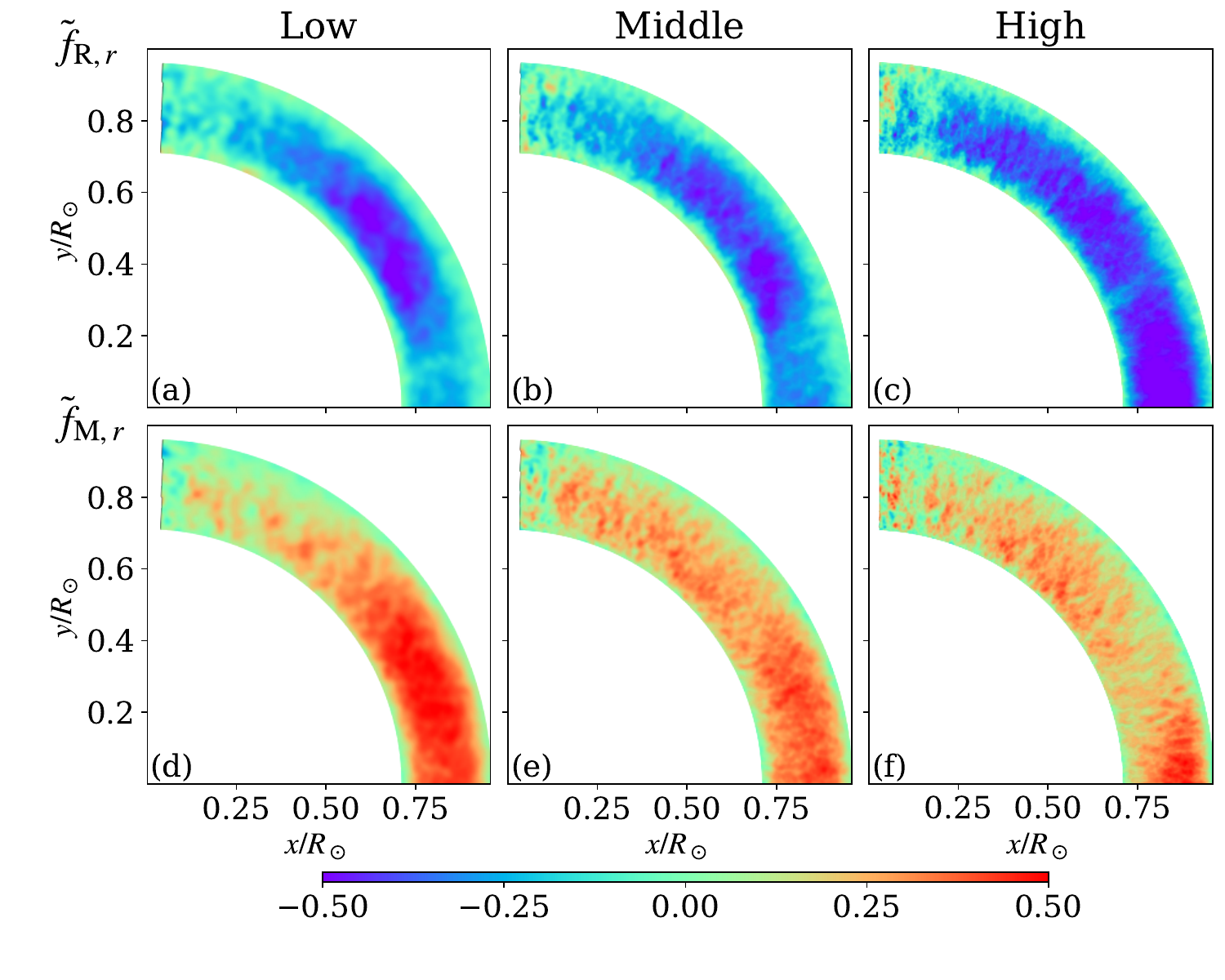}
    \caption{The same figure as Fig. \ref{normflux4_1}, but for $\tilde{f}^2_{\mathrm{R},r}$ and $\tilde{f}^2_{\mathrm{M},r}$, which correspond to the scale of $120~\mathrm{Mm}\le L_m<240~\mathrm{Mm}$ is shown.
		As resolution is increased, the negative correlation $\tilde{f}^2_{\mathrm{R},r}$ becomes stronger and the positive correlation $\tilde{f}^2_{\mathrm{M},r}$ becomes weaker.}
	\label{normflux4_2}
	\end{center}
\end{figure}

Fig. \ref{normflux4_2} shows the $\tilde{f}^2_{\mathrm{R},r}$ and $\tilde{f}^2_{\mathrm{M},r}$, which correspond to the spatial scale of $120~\mathrm{Mm}\le L_m<240~\mathrm{Mm}$.
The negative dimensionless velocity correlation $\tilde{f}^2_{\mathrm{R},r}$ is strongest in the High case and this trend also applies for  $\tilde{f}_{\mathrm{R},r}^1$. We find the absolute value of the correlation in this scale increases compared with $i=1$, i.e., $|\tilde{f}_{\mathrm{R},r}^2| > |\tilde{f}_{\mathrm{R},r}^1|$ in all the cases. On a smaller scale, the rotational influence becomes less efficient. This pattern increases the negative velocity correlation.
As far as the magnetic field $\tilde{f}^2_{\mathrm{M},r}$ is concerned, the result does not change much from $i=1$ scale.
The dependence of the magnetic correlation on the spatial scale is smaller than that of the velocity correlation.\par
As for $\tilde{f}^3_{\mathrm{R},r}$ and $\tilde{f}^3_{\mathrm{M},r}$, they are almost the same as $\tilde{f}^2_{\mathrm{R},r}$ and $\tilde{f}^2_{\mathrm{M},r}$, respectively, which are not shown here.\par
\begin{figure}
	\begin{center}
		\includegraphics[width = 0.5\textwidth]{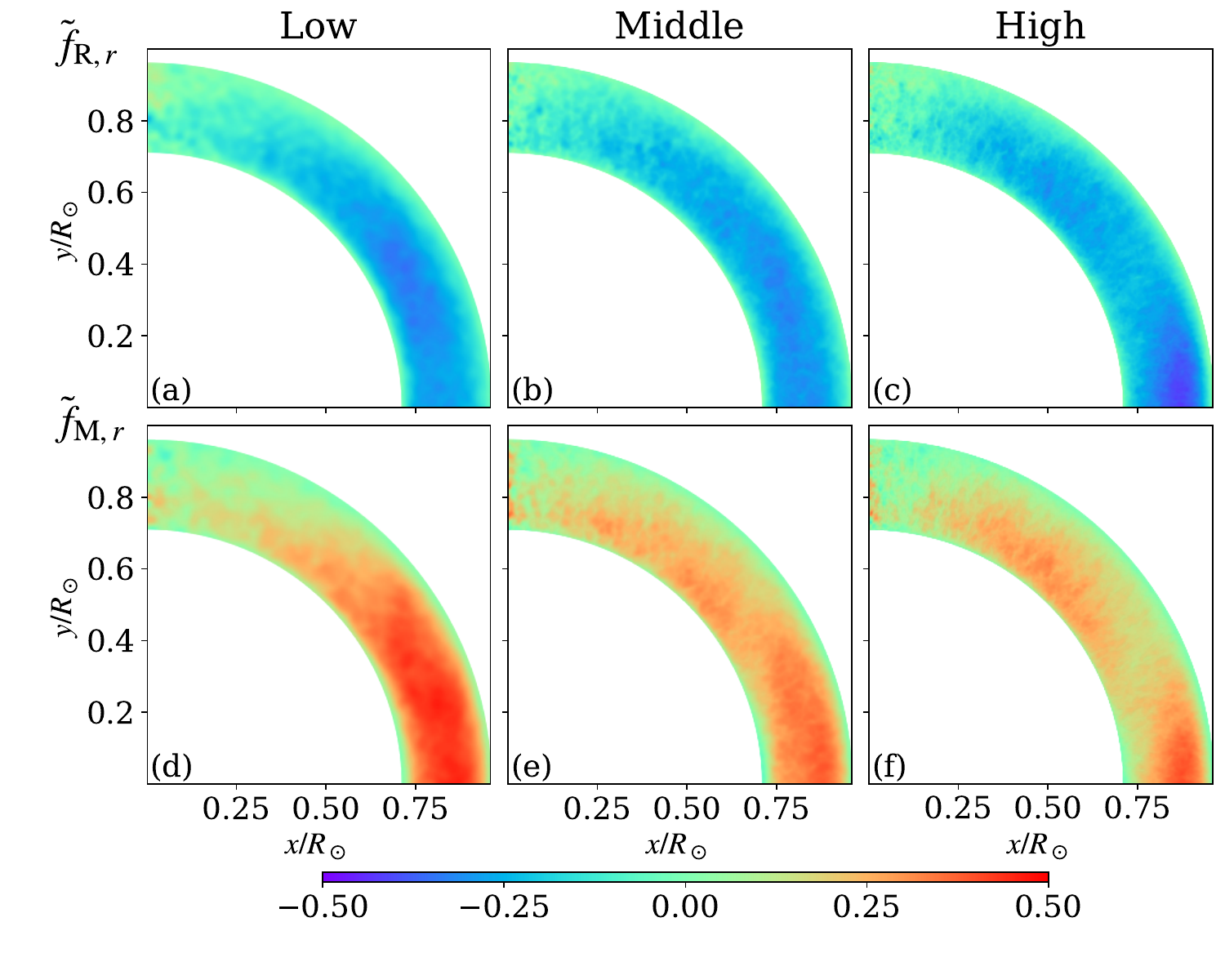}
    \caption{The same figure as Fig. \ref{normflux4_1}, but for $\tilde{f}^4_{\mathrm{R},r}$ and $\tilde{f}^4_{\mathrm{M},r}$, which correspond to the scale of $L_m<60~\mathrm{Mm}$.
		The strength of the turbulence correlation $\tilde{f}_{\mathrm{R},r}^4$ is nearly the same in all cases.}
	\label{normflux4_4}
	\end{center}
\end{figure}

The results of $\tilde{f}_{\mathrm{R}, r}^{4}$ and $\tilde{f}^4_{\mathrm{M},r}$ are displayed in Fig. \ref{normflux4_4}.
The amplitudes of the dimensionless velocity correlation $\tilde{f}^4_{\mathrm{R},r}$ are similar between the cases (Fig. \ref{normflux4_4}a, b, and c).
The dimensionless magnetic correlation $\tilde{f}^4_{\mathrm{M},r}$ is largest in the Low case. As the outward AMF $f_{\mathrm{M},r}^4$ is largest in the High case, the strength of the magnetic field is purely responsible for the strong AMT (Issue C in Section \ref{sec:introduction}).

\section{Summary}
\label{sec:conclusion}
In this work, we systematically investigate the dependence of the AMT on the spatial scale with the method suggested by MH23. We examine three simulation results in MH21, which have different resolutions: namely, Low, Middle, and High cases.
We summarize the main results of this work below:
\begin{enumerate}
	\item The absolute value of the negative dimensionless velocity correlation increases with increasing resolution, especially at the small scales.
	\item The outward Maxwell AMF is strongest at the smallest scale.
	\item The dimensionless magnetic correlation decreases with increasing resolution.
\end{enumerate}
The first result indicates that the high-resolution calculation in HK21 remains in the high Ro, while the RMS velocity is reduced in the case since the small-scale turbulence is introduced. The solar-like DR is also achieved even in the high Ro regime thanks to the magnetic AMT.\par
The second result emphasizes the small-scale magnetic field for constructing the DR. HKS22 suggested that the magnetic AMT occurs via magnetic tension. Considering that magnetic tension is most effective on a small scale, our analysis is consistent with the results of this work.\par
The third result indicates that the magnetic field becomes chaotic but strong on a small scale. Although the just chaotic magnetic field cannot transport the momentum, the remaining order with high magnetic strength can transport a large amount of angular momentum. The essential role of the small-scale magnetic field on the AMT highlights the importance of a higher resolution calculation in the future.

\section*{Acknowledgements}
H. Hotta. is supported by JSPS KAKENHI grants No. JP20K14510, JP21H04492, JP21H01124, JP21H04497, JP23H01210 and MEXT as a Programme for Promoting Researches on the Supercomputer Fugaku (JPMXP1020230504). The results were obtained using the Supercomputer Fugaku provided by the RIKEN Center for Computational Science (hp220173, hp230204, and hp230201).

\section*{Data Availability}
The analysed data underlying this article will be shared on reasonable request to the corresponding author.




\bibliographystyle{mnras}
\bibliography{diffrot_2_cite_re} 

\begin{thebibliography}{}
\makeatletter
\relax
\def\mn@urlcharsother{\let\do\@makeother \do\$\do\&\do\#\do\^\do\_\do\%\do\~}
\def\mn@doi{\begingroup\mn@urlcharsother \@ifnextchar [ {\mn@doi@}
  {\mn@doi@[]}}
\def\mn@doi@[#1]#2{\def\@tempa{#1}\ifx\@tempa\@empty \href
  {http://dx.doi.org/#2} {doi:#2}\else \href {http://dx.doi.org/#2} {#1}\fi
  \endgroup}
\def\mn@eprint#1#2{\mn@eprint@#1:#2::\@nil}
\def\mn@eprint@arXiv#1{\href {http://arxiv.org/abs/#1} {{\tt arXiv:#1}}}
\def\mn@eprint@dblp#1{\href {http://dblp.uni-trier.de/rec/bibtex/#1.xml}
  {dblp:#1}}
\def\mn@eprint@#1:#2:#3:#4\@nil{\def\@tempa {#1}\def\@tempb {#2}\def\@tempc
  {#3}\ifx \@tempc \@empty \let \@tempc \@tempb \let \@tempb \@tempa \fi \ifx
  \@tempb \@empty \def\@tempb {arXiv}\fi \@ifundefined
  {mn@eprint@\@tempb}{\@tempb:\@tempc}{\expandafter \expandafter \csname
  mn@eprint@\@tempb\endcsname \expandafter{\@tempc}}}

\bibitem[\protect\citeauthoryear{{Christensen-Dalsgaard}
  et~al.,}{{Christensen-Dalsgaard}
  et~al.}{1996}]{christensen_1996Sci...272.1286C}
{Christensen-Dalsgaard} J.,  et~al., 1996, \mn@doi [Science]
  {10.1126/science.272.5266.1286}, \href
  {https://ui.adsabs.harvard.edu/abs/1996Sci...272.1286C} {272, 1286}

\bibitem[\protect\citeauthoryear{{Gastine}, {Yadav}, {Morin}, {Reiners}  \&
  {Wicht}}{{Gastine} et~al.}{2014}]{gastine_2014MNRAS.438L..76G}
{Gastine} T.,  {Yadav} R.~K.,  {Morin} J.,  {Reiners} A.,   {Wicht} J.,  2014,
  \mn@doi [\mnras] {10.1093/mnrasl/slt162}, \href
  {https://ui.adsabs.harvard.edu/abs/2014MNRAS.438L..76G} {438, L76}

\bibitem[\protect\citeauthoryear{{Hotta} \& {Iijima}}{{Hotta} \&
  {Iijima}}{2020}]{hotta_2020MNRAS.494.2523H}
{Hotta} H.,  {Iijima} H.,  2020, \mn@doi [\mnras] {10.1093/mnras/staa844},
  \href {https://ui.adsabs.harvard.edu/abs/2020MNRAS.494.2523H} {494, 2523}

\bibitem[\protect\citeauthoryear{{Hotta} \& {Kusano}}{{Hotta} \&
  {Kusano}}{2021}]{hotta_2021NatAs...5.1100H}
{Hotta} H.,  {Kusano} K.,  2021, \mn@doi [Nature Astronomy]
  {10.1038/s41550-021-01459-0}, \href
  {https://ui.adsabs.harvard.edu/abs/2021NatAs...5.1100H} {5, 1100}

\bibitem[\protect\citeauthoryear{{Hotta}, {Rempel}, {Yokoyama}, {Iida}  \&
  {Fan}}{{Hotta} et~al.}{2012}]{hotta_2012A&A...539A..30H}
{Hotta} H.,  {Rempel} M.,  {Yokoyama} T.,  {Iida} Y.,   {Fan} Y.,  2012,
  \mn@doi [\aap] {10.1051/0004-6361/201118268}, \href
  {https://ui.adsabs.harvard.edu/abs/2012A&A...539A..30H} {539, A30}

\bibitem[\protect\citeauthoryear{{Hotta}, {Rempel}  \& {Yokoyama}}{{Hotta}
  et~al.}{2015}]{hotta_2015ApJ...798...51H}
{Hotta} H.,  {Rempel} M.,   {Yokoyama} T.,  2015, \mn@doi [\apj]
  {10.1088/0004-637X/798/1/51}, \href
  {https://ui.adsabs.harvard.edu/abs/2015ApJ...798...51H} {798, 51}

\bibitem[\protect\citeauthoryear{{Hotta}, {Iijima}  \& {Kusano}}{{Hotta}
  et~al.}{2019}]{hotta_2019SciA....5.2307H}
{Hotta} H.,  {Iijima} H.,   {Kusano} K.,  2019, \mn@doi [Science Advances]
  {10.1126/sciadv.aau2307}, \href
  {https://ui.adsabs.harvard.edu/abs/2019SciA....5.2307H} {5, 2307}

\bibitem[\protect\citeauthoryear{{Hotta}, {Kusano}  \& {Shimada}}{{Hotta}
  et~al.}{2022}]{hotta_2022ApJ...933..199H}
{Hotta} H.,  {Kusano} K.,   {Shimada} R.,  2022, \mn@doi [\apj]
  {10.3847/1538-4357/ac7395}, \href
  {https://ui.adsabs.harvard.edu/abs/2022ApJ...933..199H} {933, 199}

\bibitem[\protect\citeauthoryear{{Kageyama} \& {Sato}}{{Kageyama} \&
  {Sato}}{2004}]{kageyama_2004GGG.....5.9005K}
{Kageyama} A.,  {Sato} T.,  2004, \mn@doi [Geochemistry, Geophysics,
  Geosystems] {10.1029/2004GC000734}, \href
  {https://ui.adsabs.harvard.edu/abs/2004GGG.....5.9005K} {5, Q09005}

\bibitem[\protect\citeauthoryear{{K{\"a}pyl{\"a}}}{{K{\"a}pyl{\"a}}}{2022}]{kapyla_2022arXiv220700302K}
{K{\"a}pyl{\"a}} P.~J.,  2022, arXiv e-prints, \href
  {https://ui.adsabs.harvard.edu/abs/2022arXiv220700302K} {p. arXiv:2207.00302}

\bibitem[\protect\citeauthoryear{{Karak}, {K{\"a}pyl{\"a}}, {K{\"a}pyl{\"a}},
  {Brandenburg}, {Olspert}  \& {Pelt}}{{Karak}
  et~al.}{2015}]{karak_2015A&A...576A..26K}
{Karak} B.~B.,  {K{\"a}pyl{\"a}} P.~J.,  {K{\"a}pyl{\"a}} M.~J.,  {Brandenburg}
  A.,  {Olspert} N.,   {Pelt} J.,  2015, \mn@doi [\aap]
  {10.1051/0004-6361/201424521}, \href
  {https://ui.adsabs.harvard.edu/abs/2015A&A...576A..26K} {576, A26}

\bibitem[\protect\citeauthoryear{{Mori} \& {Hotta}}{{Mori} \&
  {Hotta}}{2023}]{mori_2023MNRAS.519.3091M}
{Mori} K.,  {Hotta} H.,  2023, \mn@doi [\mnras] {10.1093/mnras/stac3804}, \href
  {https://ui.adsabs.harvard.edu/abs/2023MNRAS.519.3091M} {519, 3091}

\bibitem[\protect\citeauthoryear{{O'Mara}, {Miesch}, {Featherstone}  \&
  {Augustson}}{{O'Mara} et~al.}{2016}]{omara_2016AdSpR..58.1475O}
{O'Mara} B.,  {Miesch} M.~S.,  {Featherstone} N.~A.,   {Augustson} K.~C.,
  2016, \mn@doi [Advances in Space Research] {10.1016/j.asr.2016.03.038}, \href
  {https://ui.adsabs.harvard.edu/abs/2016AdSpR..58.1475O} {58, 1475}

\bibitem[\protect\citeauthoryear{{Parker}}{{Parker}}{1955}]{parker_1955ApJ...122..293P}
{Parker} E.~N.,  1955, \mn@doi [\apj] {10.1086/146087}, \href
  {https://ui.adsabs.harvard.edu/abs/1955ApJ...122..293P} {122, 293}

\bibitem[\protect\citeauthoryear{{Patern{\`o}}}{{Patern{\`o}}}{2010}]{paterno_2010Ap&SS.328..269P}
{Patern{\`o}} L.,  2010, \mn@doi [\apss] {10.1007/s10509-009-0218-0}, \href
  {https://ui.adsabs.harvard.edu/abs/2010Ap&SS.328..269P} {328, 269}

\bibitem[\protect\citeauthoryear{{Schou} et~al.,}{{Schou}
  et~al.}{1998}]{schou_1998ApJ...505..390S}
{Schou} J.,  et~al., 1998, \mn@doi [\apj] {10.1086/306146}, \href
  {https://ui.adsabs.harvard.edu/abs/1998ApJ...505..390S} {505, 390}

\makeatother
\end{thebibliography}




\appendix


\bsp	
\label{lastpage}
\end{document}